\documentclass{pas}

\usepackage{amsmath}
\usepackage{booktabs}
\usepackage{xcolor}
\usepackage{graphicx}
\usepackage{dcolumn}
\usepackage{bm}
\usepackage{hyperref}
\hypersetup{colorlinks, allcolors={blue}}
\usepackage{soul}
\usepackage{siunitx}
\usepackage{multicol}

\providecommand{\araa}{ARA\&A}
\providecommand{\apj}{ApJ}
\providecommand{\apjl}{ApJL}

\providecommand{\aap}{A\&A}

\providecommand{\jcap}{J.~Cosmology Astropart.~Phys.}
\providecommand{\mnras}{MNRAS}

\providecommand{\pasa}{PASA}

\providecommand{\prd}{Phys.~Rev.~D}

\providecommand{\prl}{Phys.~Rev.~Lett.}

\providecommand{\nat}{Nature}

\usepackage[T1]{fontenc}

\usepackage{csquotes}
\usepackage{physics}
\usepackage{verbatim}
\usepackage{subcaption}
\DeclareMathAlphabet{\mathcal}{OMS}{cmsy}{m}{n}

\usepackage{multirow}

\begin{document}

\lefttitle{Publications of the Astronomical Society of Australia}
\righttitle{F. H. Panther et al. }

\jnlPage{1}{4}
\jnlDoiYr{2026}
\doival{10.1017/pasa.xxxx.xx}

\articletitt{Research Paper}

\title{Multimessenger astrophysics drives the case for an Australian Gravitational-Wave observatory}

\author{\gn{Fiona H. }\sn{Panther}$^{1, 2}$, \gn{Paul D.} \gn{Lasky}$^{2,3}$
and \gn{Nikhil} \gn{Sarin}$^{4, 5}$}

\affil{$^1$School of Physics, Mathematics and Computing, University of Western Australia, 35 Stirling Highway, Crawley 6009, Australia $^2$ARC Centre of Excellence for Gravitational Wave Discovery (OzGrav)
$^3$School of Astronomy and Physics, Monash University
$4$Kavli Institute for Cosmology Cambridge, Madingley Road, Cambridge CB3 0HA, United Kingdom
$5$Institute of Astronomy, University of Cambridge, Madingley Road, Cambridge CB3 0HA, United Kingdom}

\corresp{F. H. Panther, Email: fiona.panther@uwa.edu.au}


\begin{abstract}
The next generation of an international network of gravitational-wave observatories is now in the advanced planning stages. With Australia's substantial history of contributions in this field, its history of scientific mega-projects like the Square Kilometer Array, and its unique geographic location and conditions, it is ideally placed to host an observatory in the global network. It has already been demonstrated that a gravitational wave detector in Australia plays an important role in minimizing the sky localization uncertainty of gravitational wave events. We study a particular science case for an Australian observatory, focusing on its potential contributions to multimessenger astronomy. We demonstrate that an observatory in Australia can enable the detection of up to eight multimessenger gravitational-wave plus kilonova events per year, with localisations better than $1-10\deg^2$ in both second- and third-generation networks. We show that, in a global network of XG observatories, an Australian observatory with 2G sensitivity would still enable significant advances in multimessenger science. The full case for an Australian observatory must be considered in a much broader context; however, this work indicates it could make substantial contributions to multimessenger astrophysics. 
\end{abstract}


\maketitle
\section{Introduction}
The simultaneous observation of electromagnetic \citep[EM;][]{GW170817_multi} and gravitational-wave \citep[GW;][]{GW170817_disco} signals from the binary neutron star (BNS) merger GW170817 was a watershed moment in GW multimessenger astronomy. That such an event has not occurred in the following nine years \citep{GWTC-5} is simultaneously alarming, frustrating, and depressing. 
\citep[see][for a recent review]{Panther2026_rev}. What does it take to change this situation? 

We want to rapidly identify EM counterparts so we can ensure they are actually associated with the GW event (e.g., that they are definitely kilonovae; KNe) and not spurious associations (e.g., see claims in \cite{Moroianu2023} and subsequent analyses by \cite{Hernandez2024, Smartt2024}). This allows us to jointly analyse GW and EM data to make important inferences about, for example, the astrophysics of gamma-ray bursts and the extreme physics of neutron stars \citep[e.g.,][]{Margutti2021}. Real-time localisation also ensures that we schedule the right observations. For example, time-coincidence between serendipitous observations of an EM counterpart and single-detector GW sources can be turned into confident coincident detections through mutual information~\citep{Sarin2022, Sarin2023}. However, we cannot rely on serendipity to schedule observations that maximise scientific return.  Precise localisation also enables more precise associations with host galaxies, which enables the use of these events as `bright sirens' to measure the expansion rate of the Universe \citep[e.g.,][]{Holz2005, GW170817_cosmo}. This becomes a particularly pressing concern when searching for counterparts to more distant BNS mergers, as counterparts rapidly fade and become unobservable, even with current $8-10\,\mathrm{m}$ class telescopes \citep{Stevenson2026_kn}. 

The longer the baseline between GW observatories, the better the sky localisation one achieves, and the faster we can identify EM counterparts. A detector located in Australia provides one of the longest baselines possible between our most seasoned gravitational wave observatories in the United States and in Europe, as well as adding a unique, new North-South baseline to proposed global detector networks \citep{Wen2010, Fairhurst_2009, Fairhurst_2011, Hall2019}. Additional observatories can also provide stronger constraints on alternative theories of gravity by measuring non-GR polarisations, but we treat this as a complementary feature of additional detectors and focus here on the multi-messenger science case.

The LIGO-Virgo-KAGRA network of GW observatories~\citep{aasi15,acernese15,akutsu21} have now completed four observing runs, with near-term planned upgrades culminating in A\#~\citep{sun26}, Virgo\_nEXT~\citep{VirgoNEXT} and KAGRA-HF~\citep{akutsu25}. Those upgrades likely exhaust what current technology can do within the current facilities, prompting future plans for the US Cosmic Explorer \citep[CE; ][]{evans21} and Europe's Einstein Telescope~\citep[ET; ][]{hild11}. However, all current detectors that are planned or actively under construction are in the Northern Hemisphere. If one of the primary drivers for an expanded network is the significant sky localisation improvement, we must consider the benefits of adding a new North-South baseline. 

The different EM observatories participating in follow-up of GWs present distinct trade-offs in field-of-view, wavelength coverage, and spatial resolution, which will affect how GW follow-up is done, even with more precise localisations. The Vera C. Rubin Observatory (VRO) boasts a field-of-view of $9.6\,{\rm deg}^2$ and can observe optical KNe with a typical peak magnitude of $m_r\sim 24$ out to $\sim 400\,{\rm Mpc}$ with single 30-s exposures \citep{Rubin_facility}. However, VRO is expected to detect $\sim30$ KN `imposters' -- transients such as Type IIb supernovae that mimic KNe in photometric surveys -- per month, many of which could end up temporally and spatially associated with GW events \citep{Stevenson2026_kn}. Confirming which of these events correspond to the expected $\sim$5 kilonovae per year detectable with a signal-to-noise greater than 5 \citep{Stevenson2026_kn} will require dedicated spectroscopic follow-up, requiring well-coordinated and pre-approved programs to catch the fast-fading kilonova. 

On the other hand, space-based observatories like the Nancy Grace Roman Space Telescope offer high-resolution near-infrared (NIR) imaging, which is crucial for detecting lanthanide-rich KN emission as it rapidly shifts into NIR wavelengths, with up to 4–21 events expected to be detected during A\# \citep{Andreoni2024_NGRST}. However, Roman's smaller field-of-view ($\sim0.28\,{\rm deg}^2$) similarly requires narrow initial error boxes of $\lesssim10\deg^2$ to make mosaic tiling feasible. In both cases, the efficiency of electromagnetic follow-up is restricted by the GW localisation area. Gamma-ray burst (GRB) enabled follow-up offers an alternative pathway to localisation, separate from GW observation. However, GRB observability drops sharply outside an effective viewing angle of $\sim 15\deg$ in the local Universe~\citep{Sarin2022_linking}. Furthermore, the upcoming demise of \textit{Swift} will leave a blank space in the landscape for well-localised GRBs, although upcoming missions such as COSI~\citep{Neights2026} offer some source of optimism. 

In this Paper, we ask: \textit{what GW network sensitivity is required to see `regular' multimessenger follow-up of BNS mergers?} This vague statement takes on quantitative meaning by describing network sensitivity in terms of a) the number of GW observatories (which primarily affects the sky localisation necessary for EM follow-up) and b) the individual sensitivity of observatories in terms of their noise properties, which affects how far into the Universe each detector can see. We choose to define `regular' multimessenger observations as more than five per year, such that individual PhD students can perform quantitative population analyses within the lifetime of their degrees.

This Paper is organized as follows. In Section \ref{sec:sims}, we describe our simulation set-up, including our detector network, sensitivity curves and our computation of EM properties. In Section \ref{sec:2g}, we describe the results for our second-generation, equal sensitivity networks with and without an Australian detector. In Section \ref{sec:xg}, we discuss the introduction of an Australian detector to next-generation (XG) networks consisting of CE and ET, and investigate the impact of a detector with poorer sensitivity at the lowest frequencies. Finally, we discuss our results in comparison to other trade studies that focus on future network configurations, and describe the benefits for low-latency multimessenger astrophysics that an Australian GW detector can provide.

\section{Simulated networks and events}\label{sec:sims}
\subsection{Gravitational waves}
We simulate the performance of networks of GW observatories to infer the properties of BNS mergers. We define sets of networks with and without an observatory in Australia. In all cases, we define the network to include an observatory site in the US and an observatory site in Europe. We consider two scenarios: a second-generation scenario where all detectors have the same sensitivity, scaled relative to the nominal A$^\#$ amplitude spectral density (ASD)~\citep{sun26}; and  next-generation (XG) ASDs consistent with Cosmic Explorer~\citep{CosmicExplorer} and Einstein Telescope consisting of two L-shaped interferometers located in Northern Germany and the Mediterranean Sea respectively.

\begin{figure}
    \includegraphics[width=\linewidth]{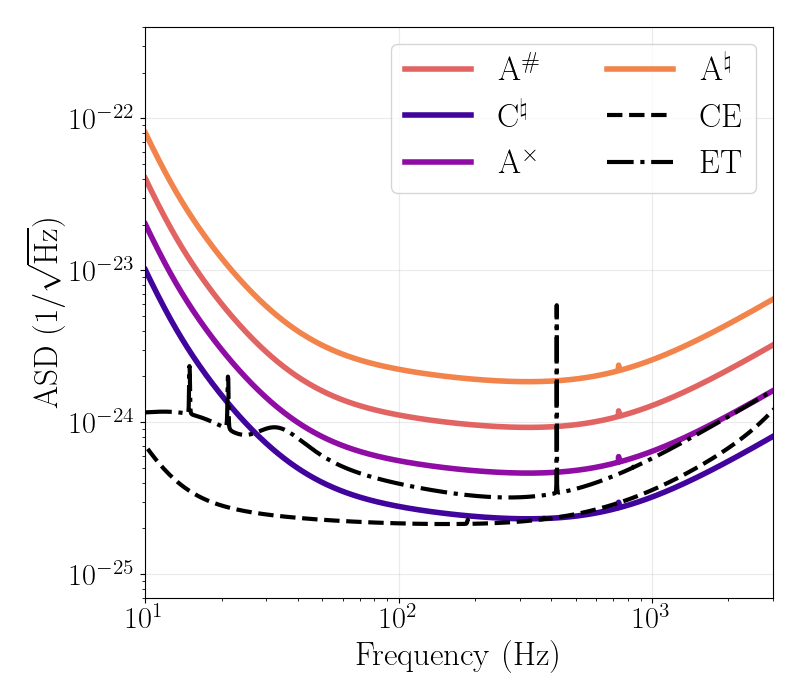}
    \caption{Detector amplitude spectral densities (ASDs) used in this work. ASD niknames are listed in Table \ref{tab:sensitivities} }
    \label{fig:ASD}
  \end{figure}

We investigate how increasing or decreasing the sensitivity of the entire network in the 2G scenario, or of the Australian observatory in the XG scenario, impacts the multi-messenger scientific output of the network. We do this by identifying multiplicative sensitivities relative to the baseline. For example, we define an ASD that we call A$^\times$ (pronounced A-double-sharp) that is twice as sensitive as A$^\#$ and is calculated by multiplying the A$^\#$ ASD by a factor of two. We list the various ASDs we use for the Australian observatory in Table~\ref{tab:sensitivities}, and show those as well as the CE and ET ASDs in Figure~\ref{fig:ASD}.

We simulate BNS mergers with the \texttt{TaylorF2} waveform model \citep{TaylorF2}, injecting 1,000 binaries into each network of observatories. We draw component masses from a Gaussian mixture model with mixing fraction 0.35 following \cite{sarin20}, where the lower peak is a normal distribution with mean and standard deviation $(\mu,\sigma)=(1.31, 0.11)M_\odot$ consistent with the Galactic neutron star mass distribution~\citep[e.g.,][]{Kiziltan2013, farrow19} and the upper peak has $(\mu,\sigma)=(1.80,0.21)M_\odot$ consistent with GW190425 \citep{GW190425} and the Galactic neutron stars in other binaries. We fix the dimensionless tidal deformability for all events $\tilde{\Lambda}=500$ and asymmetry parameter $\delta\tilde{\Lambda}=0.5$, with luminosity distances sampled uniformly in comoving volume out to a maximum redshift of $z_{\rm max}=0.2$. Our binaries are spin-aligned and drawn from a low-spin prior, $s_{iz}\in \mathcal{U}(-0.05, 0.05$). All other parameters are drawn from standard BNS priors \citep{Callister2021_thesarus}. 

We perform inference using the Fisher Information Formalism as implemented in \texttt{GWBench}~\citep{gwbench}, which allows us to avoid the computational challenge of full Bayesian inference on GW events. We only consider events with network signal-to-noise (SNR) $\rho_{\rm net}\ge15$ and where the Fisher information matrix is well conditioned; see Appendix~\ref{app:fif} for details. While we include the full 18 parameters in the Fisher matrix calculation, as we are only concerned with electromagnetic counterparts, we only consider the components corresponding to the 90\% CI of the sky localisation error ellipse, $\Omega_{90}$, and the standard deviation of the luminosity distance $d_L$ in our subsequent analysis.
\begin{table}
\centering
\begin{tabular}{llcc}
\hline
Nikname & Pronunciation & ASD improvement factor \\
\hline
A$^\natural$ & A-natural & $\times 0.5$ \\
A$^\#$ & A-sharp & $\times 1$ \\
A$^\times$ & A-double-sharp & $\times 2$ \\
C$^\natural$ & C-natural & $\times 4$\\
\hline
\end{tabular}
\caption{Summary of detector sensitivity niknames and their corresponding
scaling factors relative to the baseline second-generation, A$^\#$ PSD.}
\label{tab:sensitivities}
\end{table}

\subsection{Electromagnetic Properties}
We define a `gold' sample of GW events, which are those that can be localised to $\Omega_{90}<100\deg^2$, and can therefore be efficiently targeted by existing follow-up facilities and future, comparable facilities. This is consistent with the `gold' criterion for Rubin Target-of-Opportunity follow-up~\citep{Andreoni2024_too}. We model prompt GRB emission and KN emission on two timescales. 

We model optical/UV emission using the two-component KN model implemented in \textsc{Redback}~\citep{redback} following \citet{Sarin2024_heating}. 
For early-time emission, we calculate the peak observed r-band magnitude $m_{\rm r,peak}$ using standard filters from VRO. We describe the model and our choice of parameters in more detail in Appendix \ref{app:kn}.
We define our KN sample as those that reach $m_{\rm r,peak}<24.5$, which is the 30\,s-exposure limiting magnitude for VRO.

We also compute the magnitude of each event 30 days post-merger using JWSTs F150W filter and the same two-component KN model. We select the JWST KNe population as those that are brighter than the JWST F150W magnitude of $m_{\rm F150W}=29$, which was comparable to the observed magnitude of AT2023vfi at $\sim 29$ days post GRB230307A~\citep{Levan2024}.

We consider the capabilities of these existing facilities in light of uncertainty around the capabilities of electromagnetic facilities that will operate during the XG era to form a benchmark. The results presented in this Paper can be readily updated for future proposed facilties. 

Finally, we define events as detectable GRBs as those that have their orbital angular momentum oriented within $\iota < \pm30\deg$ of the Earth such that their relativistic jets may be observable in gamma and/or x rays \citep{Rouco2023_jetopening}. For simplicity we assume \citep[perhaps unwisely, e.g. ][]{Murguia2017_sgrbprops, Salafia2020_jetprop} that all BNSs generate such relativistic jets independent of the progenitor masses and spins. This allows us to estimate an upper limit on the rate at which BNS events with counterpart GRBs are produced in the local Universe.

\section{Second-generation networks}\label{sec:2g}
\begin{figure}
    \centering
    \includegraphics[width=1.0\linewidth]{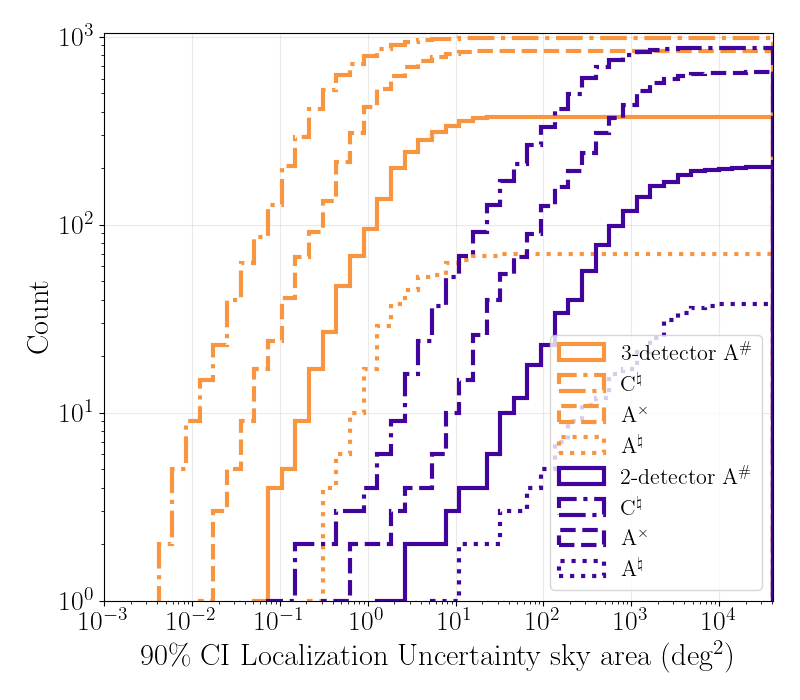}
    \caption{Cumulative distributions for the 90\% credible interval for the sky localization area ($\Omega_{90}$) for each detector network in a homogeneous network.}
    \label{fig:areas}
\end{figure}

We begin by defining homogeneous networks of GW observatories such that each observatory has the same sensitivity. Figure~\ref{fig:areas} shows the cumulative distribution of the sky localisation $\Omega_{90}$ for the recovered GW signals for our synthetic events using the various network configurations. The orange cumulative distributions show networks with an Australian observatory (i.e., a three-detector network) and the purple curves show those without (i.e., two-detector network). The different line-styles are for the different sensitivities of each observatory.

Figure~\ref{fig:areas} shows that the addition of the third observatory provides significant improvement to the median sky localisation at fixed network sensitivity, reducing it to less than a single VRO pointing ($\Omega_{90}<9.6\deg^2$) irrespective of detector sensitivity. The improvement in the total number of events observed with the addition of the third observatory is more modest, consistent with the fact that SNR adds in quadrature with the number of detectors of equal sensitivity.

\begin{figure}
    \centering
    \includegraphics[width=1.0\linewidth]{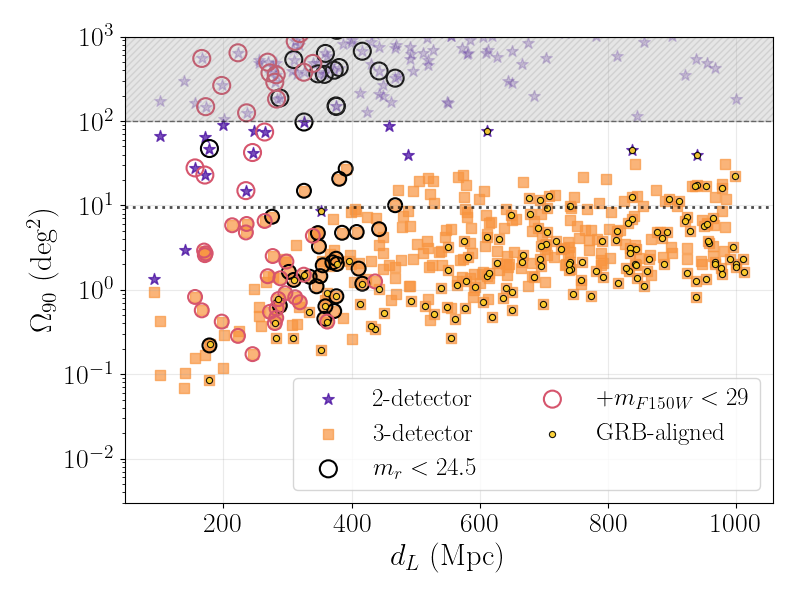}
    \caption{Comparison of injected BNS events recovered by 2-detector (purple stars) and 3-detector networks (yellow squares) at A$^\#$ sensitivity. The hatched region indicates events that fall outside of the VRO `gold' event designation. The dotted horizontal line indicates the single-pointing FoV of VRO. Events with KNe with $m_{r, peak}>24.5$ are indicated with black open circles that surround recovered injections. Events with $m_{r, peak}>24.5$ and $m_{F150}>29$ at 30 days post-explosion are indicated with pink open circles. Events that are GRB-aligned are indicated with an additional yellow circle overlaid.}
    \label{fig:asharp_summary}
\end{figure}

The localisation improvements seen in Figure~\ref{fig:areas} clearly affects multimessenger detection rates. Consider, for example, the A$^\#$ two- and three-detector homogeneous networks; Figure~\ref{fig:asharp_summary} shows the sky localisation and injected luminosity distance of all events recovered in GWs, where the symbol represents whether it is also recovered as a prompt GRB and/or a KN and/or KN afterglow. 
We summarise Figure~\ref{fig:asharp_summary} as follows:
\begin{itemize}
    \item The two-detector network detects $20.3\%$ of all injections, compared to $37.6\%$ by the three-detector network (see also the solid curve in Fig. \ref{fig:areas}). 
    \item `Gold' events: 1.9\% of all recovered events can be localized to better than $100\deg^2$ by the two-detector network, compared to $37\%$ by the three-detector network. 
    \item KN detection rate: $4.3\%$ of all injected events have $m_{\rm r,peak}<24.5$ for the two-detector network, and hence can be observed by VRO, compared to $5.5\%$ for the three-detector network. 
    \item NIR KN detection rate: $0.5\%$ (2.8\%) of all injections have $m_{\rm F150W}<27$ for the two- (three-) detector network, and can therefore be seen by JWST. 
    \item Prompt GRB emission and x-ray afterglow: the two-detector network recovers around $30\%$ of events with $\iota\leq30\deg$, compared to $\approx80\%$ for the three-detector network.
\end{itemize}

\begin{figure*}
    \centering
    \includegraphics[width=1.0\linewidth]{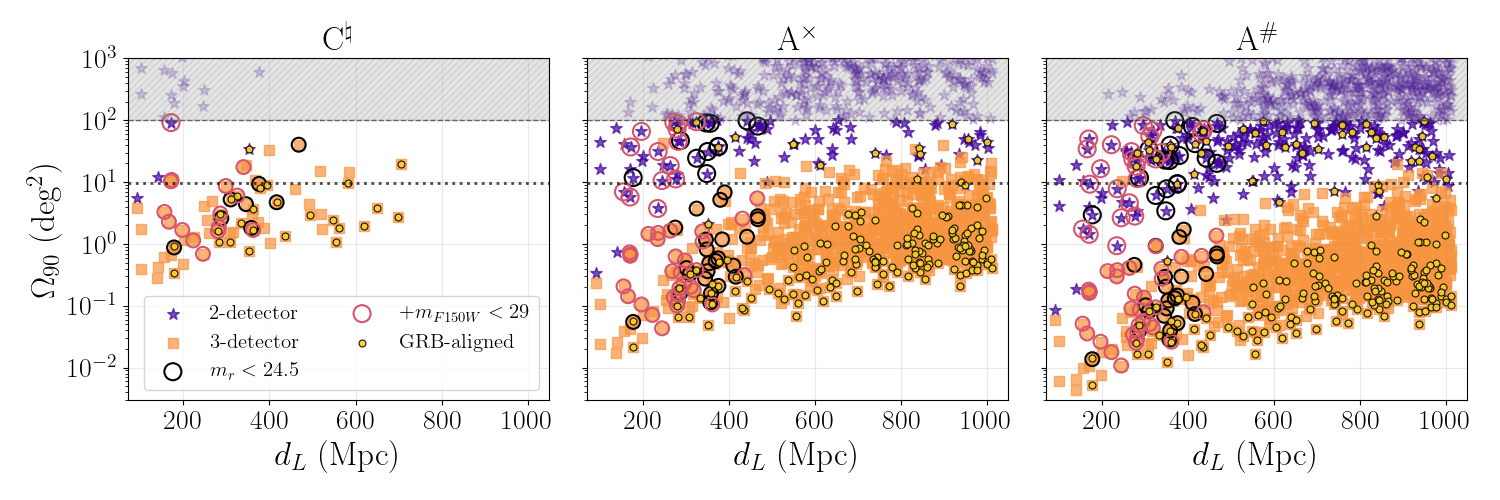}
    \caption{As Fig. \ref{fig:asharp_summary}, but for homogeneous networks with sensitivities of A$^\natural$ (left panel), A$^\times$ (middle panel), and  C$^{\natural}$ (right panel).}
    \label{fig:equal_sens}
\end{figure*}

\begin{figure*}
    \centering
    \includegraphics[width=1.0\linewidth]{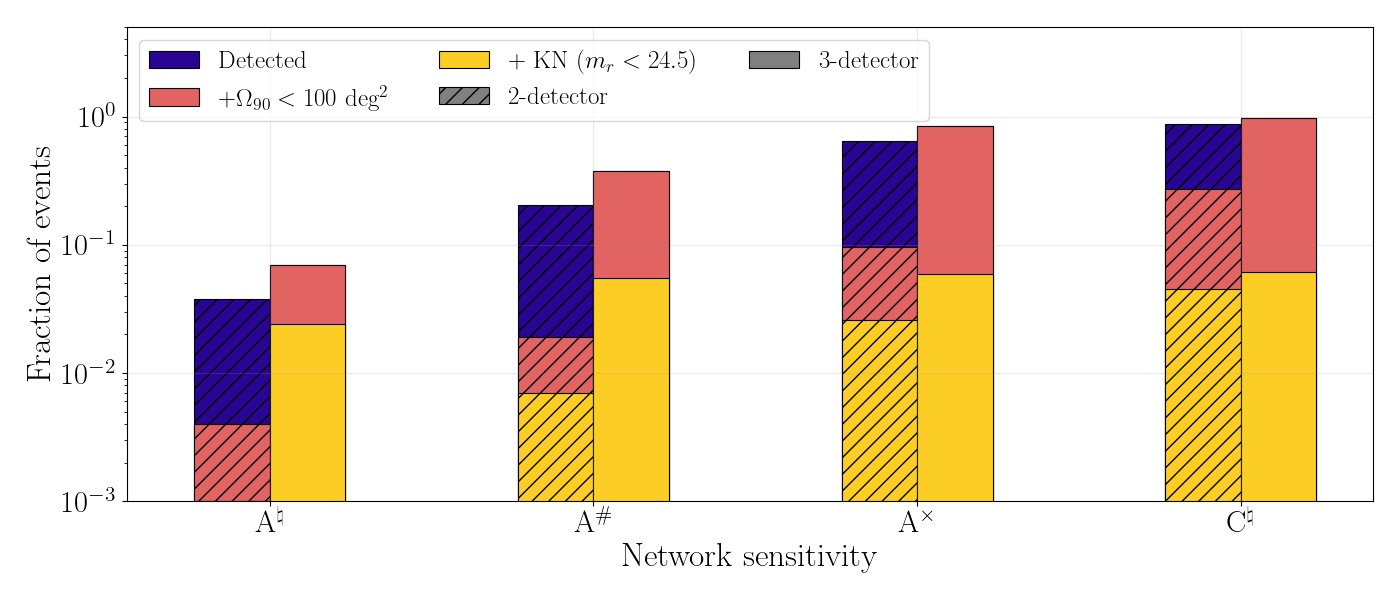}
    \caption{Cumulative fraction of events that are detected (blue), detected and localized to better than $100\deg^2$ (pink), detected and localized to better than $100\deg^2$ that produce detectable KN (yellow) for two- (hatched) and three-detector (unhatched) networks with identical sensitivity.}
    \label{fig:barsequal}
\end{figure*}

Figure~\ref{fig:equal_sens} shows the same as Figure~\ref{fig:asharp_summary} but for the homogeneous A$^\natural$ (left panel), A$^\times$ (middle), and C$^{\natural}$ (right) networks of detectors. The cumulative fraction of events that fall into different categories are shown in Figure \ref{fig:barsequal}, clearly showing the improvement in detection rates that can be achieved as detector sensitivity improves, and as we add the third, Australian detector. The main take-aways of Figures~\ref{fig:asharp_summary}, \ref{fig:equal_sens}, and \ref{fig:barsequal} are:
\begin{itemize}
    \item A$^\natural$ is most significant: the third detector has the largest impact at this lowest sensitivity, increasing the fraction of injections detected from $3.8\%$ to $7\%$, and `gold' events from $0.4\%$ to $7\%$. VRO KN-bright events increase from $1.4\%$ to $2.4\%$ of all injections, all of which are localised to better than $100\deg^2$.
    \item Sensitivity: doubling the detector sensitivity increases the median sky localisation uncertainty by a factor of $\approx2$; adding a third detector of equal sensitivity reduces median $\Omega_{90}$ by $\approx100$.
    \item GRB-aligned events: these are better localised than off-axis events at fixed luminosity distance, because the quadrupolar GW emission is strongest along the orbital axis.
    \item C$^\natural$ network: the two-detector network recovers a significant fraction of the UVOIR-bright events, but only $4.7\%$ of these can be localised within a single VRO pointing, compared to $97\%$ for the three-detector network.
\end{itemize}%

\subsection{Astrophysical event rates}
We estimate the expected rate of events satisfying each selection criterion by scaling the detected fraction of our simulated population to the observed BNS merger rate. Our population of $N_\mathrm{sim} = 1000$ injections is drawn uniformly in comoving volume $V_\mathrm{c}$ out to a maximum redshift $z_\mathrm{max} = 0.2$. The expected number of BNS mergers per year within this volume is
\begin{equation}
    N_\mathrm{exp} = \mathcal{R}\,V_\mathrm{eff}, \qquad
    V_\mathrm{eff} \equiv \int_0^{z_{\rm max}}\frac{1}{1+z}
    \frac{\mathrm{d}V_\mathrm{c}}{\mathrm{d}z}\,\mathrm{d}z ,
\end{equation}
where $\mathcal{R}$ is the BNS merger rate density that can be measured from a catalogue of detected GW events. Each injection then carries a weight
\begin{equation}
    s = \frac{N_\mathrm{exp}}{N_\mathrm{sim}}
      = \frac{\mathcal{R}\,V_\mathrm{eff}}{N_\mathrm{sim}} ,
\end{equation} 
that can be used to scale our injection recovery rate to an astrophysical rate. We adopt the GWTC-5 \texttt{fullpop} BNS merger rate density of  $\mathcal{R}= 59.3^{+95.4}_{-43.9}\,\mathrm{Gpc^{-3}\,yr^{-1}}$~\citep{gwtc5rnp}, assumed to be constant over the redshift range of interest, and propagate the 90\% CI to obtain the upper and lower bounds on the values quoted in Table \ref{tab:rates_equal}. These uncertainties reflect only the statistical uncertainty on the merger rate, as opposed to the BNS population model or KN model. These uncertainties are not expected to be significant in the context of the uncertainties on the BNS merger rate. In the case of zero recovered injections in a category, we quote an upper limit on the recovery rate.

\begin{table*}
\centering
\begin{tabular}{llcccccc}
\hline
Sensitivity & $N_\mathrm{det}$ & Detection fraction & $\Omega < 100$ deg$^2$ rate & VRO rate & VRO+JWST rate & GRB-aligned rate & True MM rate \\
\hline
A$^\natural$ & 2 & $0.04$ & $0.52^{+0.84}_{-0.38}$ & $0.13^{+0.21}_{-0.10}$ & $0.13^{+0.21}_{-0.10}$ & $0.13^{+0.21}_{-0.10}$ & $<0.38^1$\\
 & 3 & $0.07$ & $9.10^{+14.64}_{-6.74}$ & $3.12^{+5.02}_{-2.31}$ & $1.82^{+2.93}_{-1.35}$ & $2.99^{+4.81}_{-2.21}$ & $0.78^{+1.25}_{-0.58}$ \\
\hline
A$^\#$ & 2 & $0.20$ & $2.47^{+3.97}_{-1.83}$ & $0.91^{+1.46}_{-0.67}$ & $0.65^{+1.05}_{-0.48}$ & $0.52^{+0.84}_{-0.38}$ & $<0.38^1$ \\
 & 3 & $0.38$ & $48.88^{+78.64}_{-36.19}$ & $7.15^{+11.50}_{-5.29}$ & $3.64^{+5.86}_{-2.69}$ & $15.99^{+25.72}_{-11.84}$ & $1.04^{+1.67}_{-0.77}$ \\
\hline
A$^\times$ & 2 & $0.65$ & $12.48^{+20.08}_{-9.24}$ & $3.38^{+5.44}_{-2.50}$ & $1.82^{+2.93}_{-1.35}$ & $2.21^{+3.56}_{-1.64}$ & $0.39^{+0.63}_{-0.29}$ \\
 & 3 & $0.84$ & $109.72^{+176.51}_{-81.23}$ & $7.67^{+12.34}_{-5.68}$ & $4.03^{+6.48}_{-2.98}$ & $20.02^{+32.21}_{-14.82}$ & $1.04^{+1.67}_{-0.77}$ \\
\hline
C$^\natural$ & 2 & $0.87$ & $35.49^{+57.09}_{-26.27}$ & $5.85^{+9.41}_{-4.33}$ & $3.12^{+5.02}_{-2.31}$ & $6.37^{+10.25}_{-4.72}$ & $0.78^{+1.25}_{-0.58}$ \\
 & 3 & $0.98$ & $127.53^{+205.16}_{-94.41}$ & $7.93^{+12.76}_{-5.87}$ & $4.29^{+6.90}_{-3.18}$ & $20.02^{+32.21}_{-14.82}$ & $1.04^{+1.67}_{-0.77}$ \\
\hline
\end{tabular}

\caption{Expected rates of EM-bright KNe per year for each network sensitivity, scaled to the GWTC-5 BNS merger rate of $59.3^{+95.4}_{-43.9}$ Gpc$^{-3}$ yr$^{-1}$ \citep{gwtc5rnp}. The superscript $^1$ indicates the 90\% upper bound based on a single $N_\mathrm{sim}=1000$ realization (see Appendix \ref{app:fif} for details). Values that do not improve with added detectors or increased sensitivity indicate that the network is saturated and is detecting all available injections that satisfy the category criteria.}
\label{tab:rates_equal}
\end{table*}
Table \ref{tab:rates_equal} gives the event rates of BNS merger events that fall into various categories described above. We note that the sGRB event rate is over-estimated due to our extreme assumption about jets being visible for $\iota = \pm30\deg$. The GRB-aligned event rate scales as $1-\cos\theta_{\rm jet}$ for arbitrary jet opening angles. In the context of the true detection rate of events localized to better than $100\deg^2$ that can be associated with KN emission, we find that adding a third detector in Australia in the A$^\#$ era can increase the number of EM-bright GW events to a rate that would sustain multiple PhD students. A network with at least A$^\sharp$ sensitivity provides $>5$ UVOIR bright KNe per year, with several of these sufficiently luminous to be studied at 30-days post-maximum with a JWST-like telescope.

\section{Next-Generation Networks}
\label{sec:xg}

\begin{figure}
    \centering
    \includegraphics[width=1.0\linewidth]{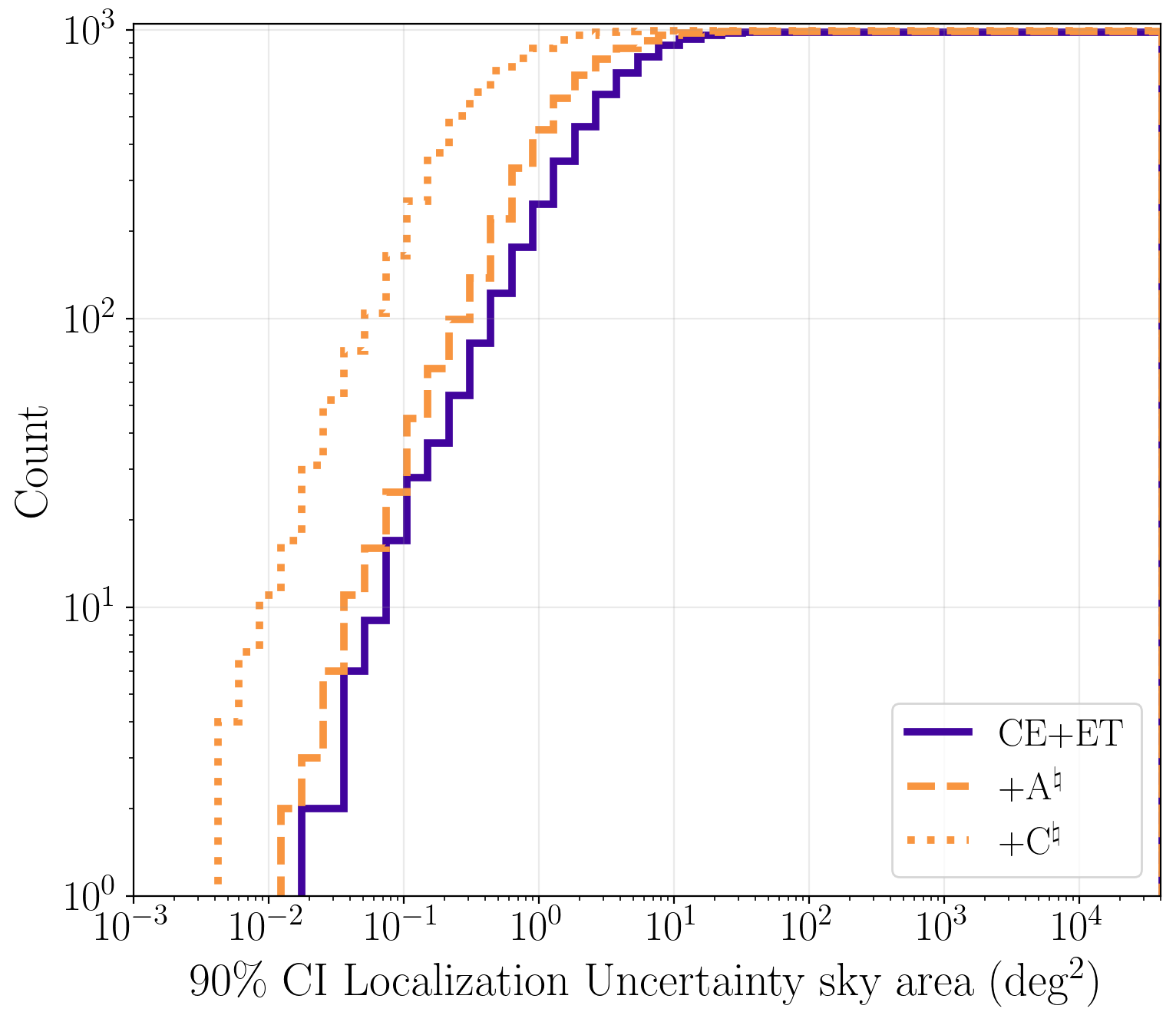}
    \caption{90\% credible interval for the sky localization area ($\Omega_{90}$) for heterogeneous detector networks that incorporate an Australian detector at A$^\natural$ (dashed) and C$^\natural$ (dotted) sensitivity, compared to the baseline CE+ET network (solid purple).}
    \label{fig:areas_xg_all}
\end{figure}

We model next-generation (XG) detector network configurations based on a single, $40\,\mathrm{km}$ L-shaped CE detector, and an ET that consists of two L-shaped detectors, one located in Northern Germany, and the other in the Mediterranean. The two ET detectors are arranged in the ``2L-$45\deg$" arrangement described in \citet{branchesi23}, optimised for multimessenger astronomy. Figure~\ref{fig:areas_xg_all} shows the cumulative count of recovered $\Omega_{90}$ for the CE+ET network (solid-purple curve), and with an Australian A$^\natural$ (dashed curve) and C$^\natural$ (dotted curve) observatory added. This shows that adding an Australian detector with either sensitivity improves the overall localization precision of the network.

Because we only inject BNS events out to $z\sim 0.2$ and our sample is volume limited, there is no substantial difference in the rate at which detector networks can detect BNS mergers. Instead, we focus on localization precision of the volume-limited sample. The key results for the XG network configurations are:
\begin{itemize}
    \item `Gold' events: all XG networks localise $\sim99\%$ of injections to better than $100\deg^2$.
    \item Single VRO pointing: the third detector increases the fraction of events localised within a single VRO pointing from $85\%$ to $98\% (99\%$) of events by including an A$^\natural$ (C$^\natural$) detector.
    \item KN detection rate: All networks recover all VRO-detectable KNe ($6\%$ of all injections), and the depth to which VRO can observe in a single 30s exposure becomes the bottleneck for KN discovery. 
    \item NIR KNe and GRBs: these networks likewise recover all VRO+JWST-detectable KNe and GRB-aligned events.
\end{itemize}

\begin{figure}
    \centering
    \includegraphics[width=1.0\linewidth]{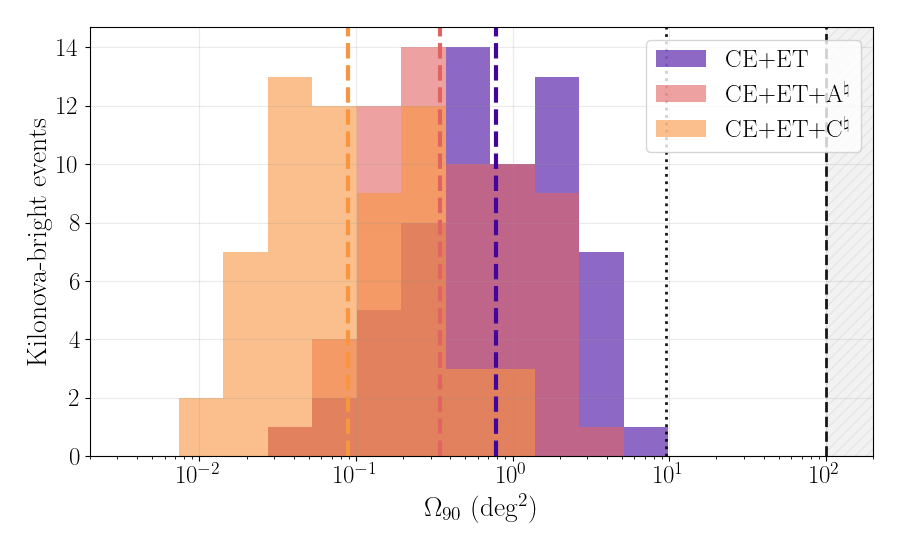}
    \caption{Distribution of recovered $\Omega_{90}$ for all KN-bright detected events by XG networks. Vertical colored dashed lines indicate median of all KN-bright detections. Dotted vertical line indicates single-pointing FoV of VRO.}
    \label{fig:knhist}
\end{figure}

The above shows that the main benefit of an additional Australian detector is to reduce the sky localization uncertainty. From Fig. \ref{fig:knhist}, the median $\Omega_{90}$ is reduced by a factor of two by the inclusion of even the least sensitive Australian detector. Adding an Australian detector allows for the majority of the KN population to be followed up with a single pointing of the Roman telescope ($\sim 0.2\deg^2$), while the CE+ET network would still require tiling of observations by the space-based observatory. 

\begin{table*}
\centering
\begin{tabular}{lcccccc}
\hline
Network & $\Omega_{90} < 100$ deg$^2$ rate & $\Omega_{90} < 9.6$ deg$^2$ rate & VRO rate & VRO+JWST rate & GRB-aligned rate & True MM rate \\
\hline
CE+ET & $128^{+205}_{-95}$ & $111^{+178}_{-82}$ & $7.9^{+13}_{-5.9}$ & $4.3^{+6.9}_{-3.2}$ & $19^{+30}_{-14}$ & $1.0^{+1.7}_{-0.77}$ \\
A$^\natural$ CE+ET+AUS & $128^{+206}_{-95}$ & $123^{+197}_{-91}$ & $7.9^{+13}_{-5.9}$ & $4.3^{+6.9}_{-3.2}$ & $19^{+31}_{-14}$ & $1.0^{+1.7}_{-0.77}$ \\
C$^\natural$ CE+ET+AUS & $129^{+207}_{-95}$ & $129^{+207}_{-95}$ & $7.9^{+13}_{-5.9}$ & $4.3^{+6.9}_{-3.2}$ & $20^{+32}_{-15}$ & $1.0^{+1.7}_{-0.77}$ \\
\hline
\end{tabular}
\caption{Astrophysical rates ($\mathrm{event\,yr^{-1}}$) of detected BNS events in a volume-limited injection set out to $z\sim 0.2$ satisfying our various cuts, scaled to the GWTC-5 BNS merger rate. The main advantage offered by the Australian detector is the increase in the number of events that can be localized to better than a single VRO pointing. }
\label{tab:rates_xg}
\end{table*}

VRO is expected to detect up to 10,000 transients per night \citep{Rubin_facility}. The future of GW follow-up may be \textit{confusion limited} if several viable transients are identified in the localization volume \citep{Stevenson2026_kn}. The most distant of these transients may quickly fade below the detection limit if they are at luminosity distances $\gtrsim 400\,\mathrm{Mpc}$, implying we may not have sufficient time to provide reliable spectroscopic classifications from even 8-m class facilities. This may result in multiple putative hosts with different redshifts being identified for a single BNS merger. One way to overcome this is to reduce the volume in which GW events are localized sufficiently that we can perform individual galaxy localizations. This will also have important benefits for dark siren cosmology as it will reduce the number of host redshifts over which we must marginalise when calculating $H_0$.

We expect that BNS progenitors (and by extension, binary black hole progenitors of dark siren sources) will trace stellar mass, and so we consider the low-redshift number density of galaxies with $M_\star>10^{10}\,\mathrm{M_\odot}$; \citet{moustakas13} gives us 
\begin{equation}
    n_\star(M_\star>10^{10}\,\mathrm{M_\odot}) \sim 5\times10^{-3}\,\mathrm{Mpc^{-3}}.
\end{equation}
We use our uncertainty estimates $\sigma_{d_L}$ for each event in our bright siren sample. This sample requires
\begin{itemize}
    \item a GW event in our injected population with a localization uncertainty of $\Omega_{90}<100\,\deg^2$,
    \item a KN counterpart with $m_\mathrm{r,\,peak}<24.5$, and 
    \item the one-sigma uncertainty in $d_L$ does not exceed $d_L$, i.e. $\sigma_{d_L}<d_L$
\end{itemize}
We calculate the 90\% CI of the posterior on $d_L$ ($\Delta d_L \equiv 1.645\,\sigma_{d_L}$) to determine the 90\% co-moving search volume of the spherical shell of width $d_L \pm 1.645\,\sigma_{d_L}$ centered on the true luminosity distance of the source\footnote{for localizations as small as those in the XG network study, we find that the covariance between parameters is negligible, justifying the shell approximation.}, i.e.  
\begin{equation}\label{eq:volume}
    V_{90} \simeq \frac{\Omega_{90}}{41253\,\mathrm{deg}^2}\,
    \frac{4\pi}{3}\,\frac{\left(d_L + \Delta d_L\right)^3
    - \left(d_L - \Delta d_L\right)^3}{(1+z)^{3}}.
\end{equation}

\begin{figure}
    \centering
    \includegraphics[width=1.0\linewidth]{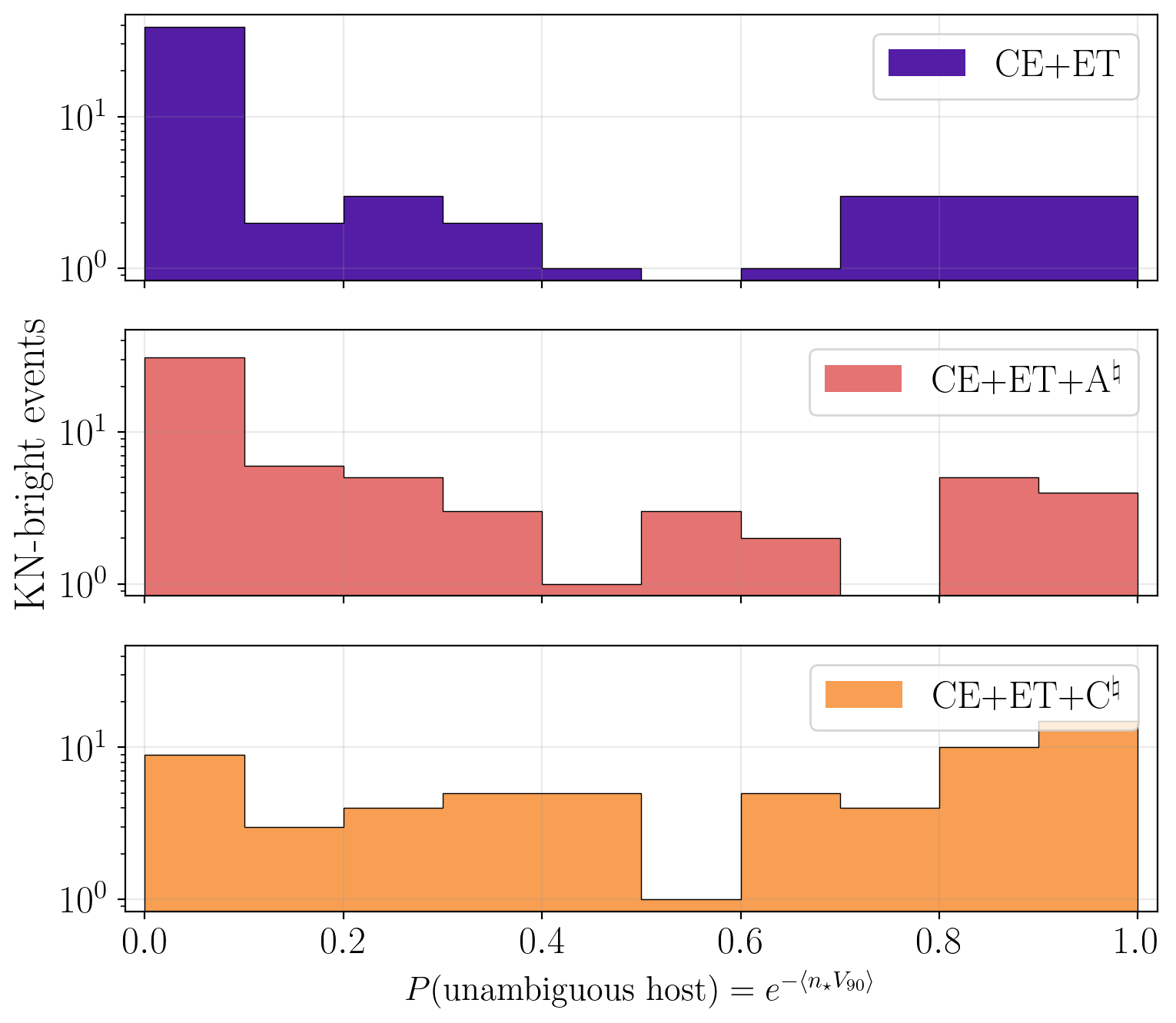}
    \caption{Probability that KN-bright events can be localized such that there is exactly one $10^{10}\,\mathrm{M_\odot}$ host galaxy (assuming uniform galaxy densities) in $V_{90}$.}
    \label{fig:p_onehost}
\end{figure}

We calculate the expected number of galaxies of randomly distributed galaxies $\lambda = \langle n_*V_{90}\rangle$ in each 90\% localization volume under the assumption that $V_{90}$ is centered on the true host for our XG networks. In Figure \ref{fig:p_onehost}, we show the probability that our bright siren population has an unambiguous host for each XG detector configuration. In a CE+ET two-site network, the median probability of finding only one host per $V_{90}$ is $P = 0.006$. Adding an Australian detector with A$^\natural$ (C$^\natural$) sensitivity increases the median probability of finding only one galaxy per $V_{90}$ to $P = 0.05$ ($P=0.65)$. 

To consider how luminosity distance affects these results, We split the recovered injections into bins of width $50\,\mathrm{Mpc}$ based on their injected $d_L$ to empirically determine the median $\sigma_{d_L}$ as a function of $d_L$. We use this to compute the expected $\Omega_{90}$ in the limit that $\langle n_\star V\rangle \simeq 1\,\mathrm{galaxy}$. This corresponds the maximum $\Omega_{90}$ for a GW event such that a one-to-one association with a host galaxy can be made from the GW data alone. 

\begin{figure}
    \centering
    \includegraphics[width=1.0\linewidth]{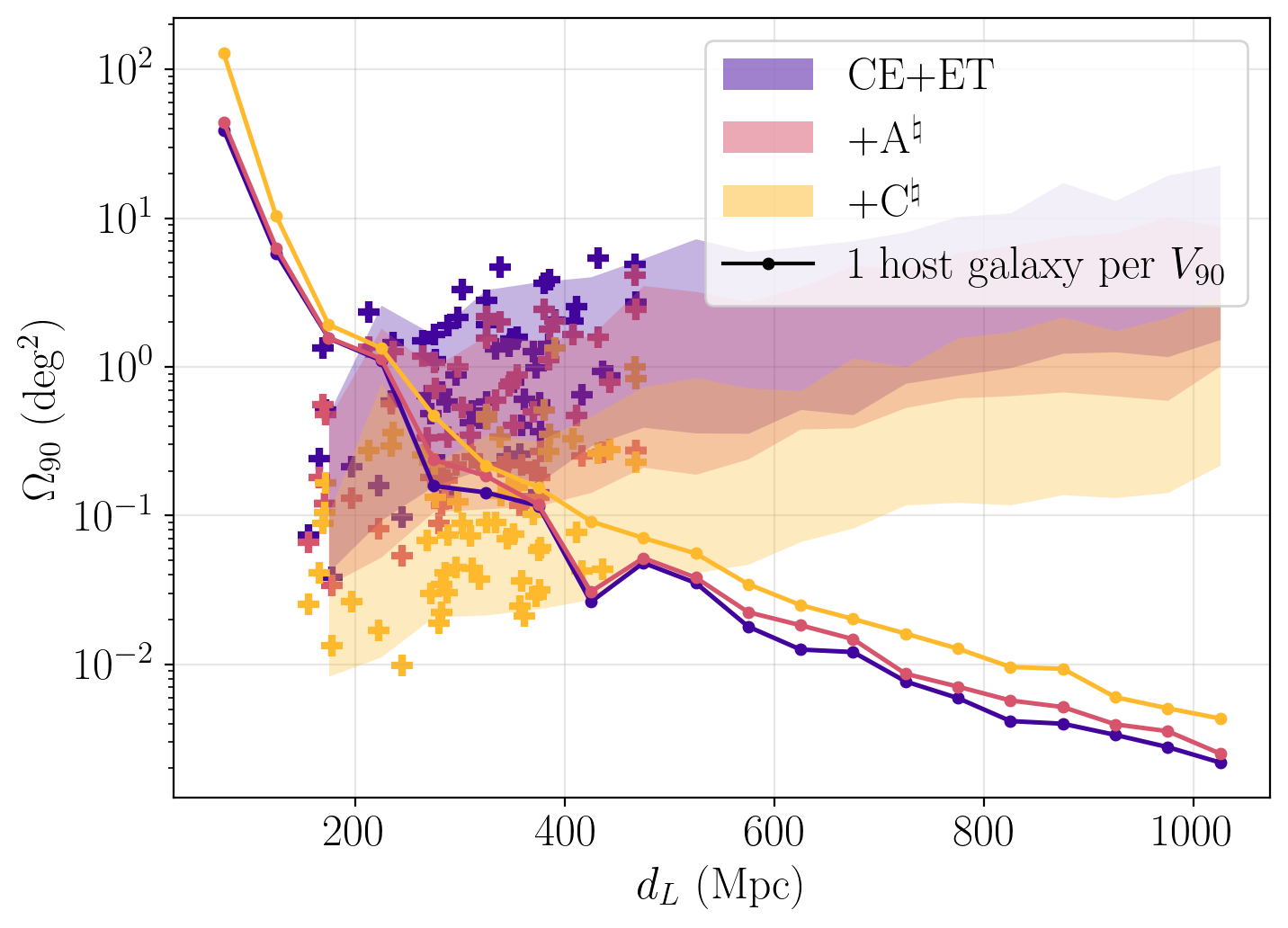}
    \caption{As above, for the XG networks Shaded bands show the 90\% CI of recovered $\Omega_{90}$ for all GW events. Scatter points show the injected luminosity distance and recovered $\Omega_90$ for KN bright events in the two-site (purple) and three-site (pink: +A$^\natural$, and yellow: +C$^\natural$) networks. Lines indicate $\Omega_{90}$ such that events can be localized to one host galaxy from gravitational wave data alone, assuming a GW $\sigma_{d_L}$ that is taken from the median in each $d_L$ bin.}
    \label{fig:confusion_xg}
\end{figure}

In the XG networks, adding an Australian detector increases the maximum luminosity distance at which KNe can be associated with a single host galaxy, as well as increasing the overall fraction of KNe with single-host associations (yellow/pink crosses below yellow/pink solid line, Fig. \ref{fig:confusion_xg}). This proves to be the most significant argument for including an Australian detector under the assumption of a 100\% duty cycle for all detectors (see Discussion section), as with a three-detector network 24\% and 61\% of all injected KNe can be localized to individual host galaxies from GW data alone for the least and most sensitive Australian detector, respectively. 

These results demonstrate that a detector network that maximises multimessenger science can also maximise our ability to do precision cosmology with GWs in coming decades. In particular, an Australian detector in an XG network, when combined with future spectroscopic galaxy catalogues that are complete to $\sim 300\,\mathrm{Mpc}$ could entirely eliminate the need to perform photometric tiling of GW localizations for KN follow-up. 

\subsection{Optimising an Australian configuration}
Two metrics to optimise when thinking about an Australian observatory in a global network are 1) the ability to do multimessenger science and 2) cost. If we simultaneously maximise multimessenger science and minimise cost\footnote{We work under the premise that cost scales with sensitivity. We hypothesise that in Australia, the relevant cost scaling is with the number of times the government changes during construction.}, we find that an Australian detector does not necessarily need to match the sensitivity of the international network. For example, the $\mathrm{CE+ET+A^\natural}$ network can localize all GW events accompanied by KNe that are detectable by VRO to $\Omega_{90}<100\deg^2$, with the vast majority having localization areas that are smaller than a single VRO pointing. Increasing the sensitivity of the Australian detector increases the fraction of events that can be localized to better than $0.1\deg^2$. However, with wide FoV UVOIR follow-up facilities, such precise localizations do not necessarily result in a greater number of discoveries. In summary, in the XG era, incorporating the `cheapest' plausible Australian detector, with a sensitivity comparable to LIGO's proposed O5/A+ observing run (i.e., similar sensitivity to our A$^\natural$), captures all multimessenger events we simulate. Sensitivity improvements, which require reduction of seismic and quantum noise levels through advanced technologies, offer only marginal benefit to the localization of the closest multimessenger events. A GW network consisting of $\mathrm{CE+ET+A^\natural}$ can detect $7.9^{+13}_{-5.9}$ KNe each year with a median localization of $\sim 0.3\deg^2$, based on the two-component KN model used in this work and scaling our recovery rates to the GWTC-5 BNS rate. 

\section{Discussion and Conclusions}

Our focus on an Australian GW detector is motivated by the strength of of the GW astronomy community in Australia, and the fact that the benefit of an Australian detector has already been demonstrated. It has been shown by \cite{Wen2010, Fairhurst_2009, Fairhurst_2011} and \cite{Hall2019} that the minimum localization area that can be resolved by a GW detector network depends on the area spanned by the detector baselines projected perpendicular to the source direction. Australia is close to antipodal to the US sites and far from Europe, so it gives the longest available baselines and the largest projected areas available to a US–Europe network. Based on the results of our 2G simulations, we find that while a 3-detector network at A$^\sharp$ sensitivity doubles the total fraction of events that can be recovered, the rate of events that are localized to better than $100\deg^2$ increses by a factor of 20. This drastically increases the size of the VRO `gold' event sample that can be followed up in low-latency, and specifically increases the rate of kilonova discovery. 

This is especially important for the more distant KNe in our sample, which would rapidly fade below the detection limit of VRO and any $8\,\rm{m}$-class facilities used for spectroscopic follow-up. We also find that the inclusion of the third site breaks the distance-inclination degeneracy that dominates two-site error in $d_L$, which is essential for precise standard siren cosmology. This would complement the observing capabilities of the Square Kilometer Array, which can also be used to further constrain the inclination angle of BNS sources through high-resolution radio imaging of jet emission \citep[e.g.][]{Gourdji2026} for sources beyond the reaches of UVOIR facilites. In addition, the substantially smaller localisation volumes allow single-host association from GW data and spectroscopic galaxy catalogs alone, especially if peculiar velocities for catalogued galaxies are known \citep{Blake:2025etn}. This could enable faster follow-up with spectroscopic facilities with very narrow fields of view. Smaller localisation volumes also reduces uncertainties contributed by the host-redshift marginalisation for dark siren cosmology. This will be complemented by current and future HI galaxy surveys \citep{Abdalla:2009wr}. There will not be an exact correspondence between the localization precision achievable for much longer-duration BNS signals studied here compared to binary black holes. Nevertheless, the reduction in the localization volume offered by a third detector could accelerate the rate at which dark sirens can achieve the $\lesssim$2\% precision required to break the Hubble tension. 

Adding an A$^\natural$ sensitivity (comparable to O5 sensitivity) Australian detector captures essentially all simulated multimessenger events in both our 2G and XG network simulations. In the XG era, it increases the probability that a proportion of these can be localized to an individual host galaxy from GW alone -- we find that the low-latency KN identification rate in the XG era is limited by the depth to which telescopes can observe if we have a 3-detector network. 

In our simulations, we assume a 100\% duty cycle for all detectors. In reality, observatories experience significant downtime, implying some multimessenger science will require single-observatory detection, which in turn relies on detection of the prompt high-energy emission~\citep{Sarin2023,jamsin26}. A third detector significantly mitigates this problem by reducing the time only a single observatory is on-sky.


Finally, we have restricted our low-latency follow-up of multimessenger sources to exisiting known EM telescope magnitude limits; our KN rate saturates because we can only see them out to $\lesssim$500 Mpc. However, in the future, new, larger ELTs will be able to detect things that are much further away. This is when having a third detector that is optimally orientied is a true advantage, as these events will always have intrinsically larger localization uncertainties.

In this Paper, we demonstrate that multimessenger astronomy, and one of its downstream beneficiary standard siren cosmology, can drive the science case for an Australian GW detector. Such a detector need not necessarily be as sensitive as future planned detectors -- adding a detector in Australia with sensitivity comparable to LIGO O5 can increase the rate of GW+KN detections wtih sky localizations $<100\deg^2$ rate from effectively zero with a two-detector network to $\sim 3$ events per year during the second-generation detector era. In the XG era, an O5 sensitivity detector can reduce the median sky localization area of KN-bright GW events to $0.1\deg^2$ -- a single pointing of the Nancy Grace Roman Space Telescope. 

\section{Acknowledgements}
The authors thank Carl Blair, David Blair, Chiara di Fronzo, Bram Slagmolen, Lilli Sun and Sheon Chua for useful discussions. FHP is supported by a Forrest Research Foundation Fellowship. This work was supported through Australian Research Council (ARC) Centre of Excellence CE230100016, Discovery Projects DP220101610 and DP230103088, and LIEF Project LE210100002. NS acknowledges support from the Kavli Foundation.

\appendix

\section{Statistical formalism}
\label{app:fif}
In the presence of an astrophysical signal, the data measured in a single gravitational wave detector is
 \begin{equation}
     d(t, \vec{\theta}) = h(t, \vec{\theta}) + n(t)
 \end{equation}
 and the probability we measure data $d$ given the parameters $\vec{\theta}$ is given by the Whittle likelihood
 \begin{equation}
     \log\mathcal{L}(d|\vec{\theta}) \propto \langle d - h(\theta), d- h(\theta)\rangle
 \end{equation}
 were we have dropped the explicit time dependence of the signal for brevity. $\langle a, b\rangle$ denotes the usual noise-weighted inner product
 \begin{equation}
    \langle a, b\rangle = 4\mathrm{Re}\int_0^{\infty}\frac{\tilde{a}(f)\tilde{b}^*(f)}{S_n(|f|)}df
    \label{eq:inner}
 \end{equation}
 where an over-tilde indicates the Fourier transform of the function, and $S_n(|f|)$ is the one-sided noise power spectral density \citep{Cutler1994}. The optimal signal-to-noise ratio (SNR, $\rho$) of a signal $h$ is then
 \begin{equation}
     \rho^2 = \langle{h, h}\rangle.
 \end{equation}
For the gravitational wave likelihood, the Fisher Information Matrix (FIM) elements are \citep[e.g. see][]{Vallisneri2008}
 \begin{equation}
     \Gamma_{ij} = \bigg\langle\frac{\partial h(t, \vec{\theta})}{\partial \theta_i},  \frac{\partial h(t, \vec{\theta})}{\partial \theta_i}\bigg\rangle,
     \label{eq:fisher}
 \end{equation}
where $i,j$ index over $\vec{\theta}$.

The reliability of the results of our simulation depends strongly on the signal-to-noise ratio for the CBC signal being sufficiently large that the FIM is not ill-conditioned. Because these ill-conditioned matricies cannot be inverted, the estimated uncertainties are not reliable. For this reason, any injection that yields a covariance matrix with condition number $c = e_\mathrm{max}/e_\mathrm{min}>10^{15}$ is excluded, where $e_{x}$ are the eigenvalues of the full covariance matrix $\Gamma^{-1}$. We choose this condition on the basis that when $c$ exceeds the inverse machine precision they cannot be numerically inverted. The condition number can be extracted directly from the \texttt{network} object supplied by \texttt{gwbench}. Any events with an $\rho_\mathrm{net}<15$ are also removed \citep{Vallisneri2008}. More recent works such as \citet{Rodriguez2013} suggest that the FIF method is unreliable for events below $\rho_\mathrm{net}<25$, however we find that most of our signals with $\rho<25$ have already been rejected on the basis of their Fisher matrices being ill-conditioned. 

We place a lower bound on the evaulation of the FIM (Eq. \ref{eq:fisher}) of $10\,\mathrm{Hz}$. The upper limit of the integral in Equation \ref{eq:inner} is $1024\,\mathrm{Hz}$ and the limit is chosen for computational efficiency. This is comparable to the merger frequency of a neutron star binary,  
\begin{equation}
    f_\mathrm{ISCO} = \bigg(6^{3/2}2\pi M_\mathrm{tot}\bigg),\quad c = G = 1.
\end{equation}
within a few percent. A small portion of the signal is lost by limiting the maximum frequency to reduce overall computation time, however the majority of the BNS SNR is accumulated at frequencies below $1000\,\mathrm{Hz}$. For a small fraction of our signals, this overestimates the sky localization precision and they are \textit{worse} by a factor of $\leq 10\%$. Limiting the lower bound on the integral in Equation \ref{eq:inner} to $10\,\mathrm{Hz}$ when evaluating Equation \ref{eq:fisher} also has the practical advantage of not requiring us to consider the rotation of the Earth without substantially decreasing the overall recovered SNR, although our sky localization areas represent a pessimistic limit.

To calculate the upper limit on the event rate in the case of non-detection, we consider the probability of recovering none of our $N_\mathrm{sim}$ injections in a given network plus injection set realization is $(1-p)^{N_\mathrm{sim}}$. Here, $p = f_\mathrm{rec}$ i.e. the recovery fraction in a given category (in this case, the `true MM rate' of events where $\Omega_{90}<100\deg^2 + \mathrm{KN} + \mathrm{GRB}$). We place a 95\% upper limit on $f_\mathrm{True\,MM} <0.003$, which corresponds to a rate of 
\begin{equation}
    \mathcal{R_\mathrm{True\, MM}}<0.38\,\mathrm{yr^{-1}}
\end{equation}.

\section{Kilonova modelling}\label{app:kn}
The assumptions made to model the KN lightcurves with the \texttt{Redback} KN model are shown in Table \ref{tab:knparams}. Ejecta masses and the dynamical ejecta velocity are not free parameters: they are computed from the component masses, tidal deformabilities, $M_\mathrm{TOV}$ and $\zeta$ using the relations of \citet{Coughlin2019}.

\begin{table}
\centering
\begin{tabular}{llp{0.4\linewidth}}
\hline
Parameter & Value & Description \\
\hline
$\zeta$ & $0.1$ & Fraction of the disk that becomes unbound \\
$M_\mathrm{TOV}$ & $2.2\,M_\odot$ & Tolman-Oppenheimer-Volkoff maximum mass \\
$v_{\mathrm{ej},2}$ & $0.05\,c$ & Disk wind ejecta velocity \\
$\kappa_1$ & $10\,\mathrm{cm^2\,g^{-1}}$ & Grey opacity, dynamical component \\
$\kappa_2$ & $35\,\mathrm{cm^2\,g^{-1}}$ & Grey opacity, disk wind component \\
$T_{\mathrm{floor},1}$ & $1000\,\mathrm{K}$ & Floor temperature, dynamical component \\
$T_{\mathrm{floor},2}$ & $1000\,\mathrm{K}$ & Floor temperature, disk wind component \\
$\tilde{\Lambda}$ & $500$ & Mass-weighted tidal deformability of the binary \\
$\delta\tilde{\Lambda}$ & $0.5$ & Tidal deformability asymmetry \\
\hline
\end{tabular}
\caption{Parameters of the two-component kilonova model
(\texttt{two\_component\_bns\_ejecta\_relation}) implemented in \texttt{redback}.}
\label{tab:knparams}
\end{table}

\section*{Data Availability}
\bibliographystyle{apsrev4-1}
\bibliography{bibliography}

\end{document}